\documentclass[prl,aps,amsfonts,amssymb,floatfix,twocolumn,superscriptaddress]{revtex4-2}

\usepackage[T1]{fontenc}
\usepackage[utf8]{inputenc}
\usepackage{enumerate}
\usepackage[toc,page]{appendix}
\usepackage{graphicx}
\usepackage{amsmath}
\usepackage{amsfonts}
\usepackage{amssymb}
\usepackage{mathtools}
\usepackage{multirow}
\usepackage{bm,hyperref,cleveref,xcolor}
\let \vec \mathbf
\newcommand{\uvec}[1]{\vec{#1}}
\newcommand{\dd}{\mathrm{d}}
\newcommand{\DD}{\mathrm{D}}
\newcommand{\ph}[1]{\mathsf{#1}}
\newcommand{\ii}{\mathrm{i}}
\let\epsilon\varepsilon
\begin{document}
\title{The active Boltzmann equation: From dilute collisions to exact kinetic theory}
\author{Thomas Ihle}
\affiliation{Institute for Physics,
Greifswald University, Greifswald, Germany}
\author{Horst-Holger Boltz}
\affiliation{Institute for Physics,
Greifswald University, Greifswald, Germany}
\author{R{\"u}diger K{\"u}rsten}
\affiliation{Institute for Physics,
Greifswald University, Greifswald, Germany}
\affiliation{Internationales Universitätszentrum, TU Bergakademie Freiberg, Freiberg, Germany}
\author{Benjamin Lindner}
\affiliation{Institute for Physics,
Humboldt University, Berlin, Germany}

\begin{abstract}
We construct an exact kinetic theory for dilute aligning self-propelled particles by closing the BBGKY hierarchy non-locally, retaining the pair correlations during each collision. Mean-field closures miss these correlations entirely and predict a divergent self-diffusion coefficient. Our theory resolves this failure and agrees quantitatively with simulations. For aligning couplings, a physical regularisation eliminates secular terms from diverging collision times. The theory yields effective Langevin equations  with analytic noise correlations which is extrapolated to higher densities.
\end{abstract}

\maketitle

Collections of self-propelled particles (SPPs) provide the most common realisation of
active matter and have been extensively studied as minimal representations of many
living and synthetic systems
\cite{vicsek_12,marchetti_13,menzel_15,baer2019,chate_20,tevrugt2025,tevrugt2026}.
Intriguing collective phenomena, including wave formation and mesoscale turbulence, can
be obtained from simplistic microscopic models \cite{bechinger_16}. One prominent class
of such models is characterised by a velocity-alignment rule among neighbouring
particles and goes back to the famous Vicsek model \cite{vicsek_95,czirok_97,nagy_07}.
Since a global theoretical framework is missing for such far-from-equilibrium systems,
many works rely on agent-based computer simulations and theories
derived by means of mean-field assumptions
\cite{bussemaker_97,bertin_06,peruani_08,bertin2009,ihle_11,chou_12,menzel_12,roman_12a,roman_12b,grossmann_13,degond2013,romensky_14,peshkov2014,reinken_18,peruani2021,benvegnen_22,solon2022,bonilla2023,kreienkamp2024,duan2025,lier2026,lardet2026,myin2026,burekovic2026}.

Such closures rest on \emph{molecular chaos}: each particle is taken to respond to the mean
field of many uncorrelated partners, so that the interaction is effectively smooth and
pair correlations may be discarded. The opposite limit is dilution. There a particle
interacts with at most one partner at a time, and the interaction it experiences is not a
smooth average but fluctuating: a sequence of discrete, possibly strong scattering events. 

Kinetic equations of Boltzmann type have been applied to aligning active matter with
considerable success~\cite{bertin_06,bertin2009,peshkov2014}. There, a collision
integral is constructed directly for the one-particle distribution from physically
motivated scattering rules, and correlations are discarded. Our starting point is different. We show how to derive the
collision integral from the microscopic equations of motion through a non-local closure
of the BBGKY hierarchy, retaining the pair correlations that a collision itself builds
up. Molecular chaos thus becomes a result to be checked rather than an input, and we
find that it fails dramatically and unphysically in a workhorse model. As an example, we show that mean-field theory cannot capture self-mixing and predicts a self-diffusion coefficient that is diverging in the deterministic limit.

In this Letter we construct this theory for anti-aligning couplings with finite collision times, show that it reproduces the relaxation of the angular mode hierarchy quantitatively where mean-field theory fails qualitatively, and extend it to aligning couplings, where diverging collision times require a physical
regularisation by quenched speed disorder. As an application we derive an effective
one-particle Langevin equation, with analytic expressions for its noise correlations and
the self-diffusion coefficient. Full derivations and extended results are presented in a
companion paper~\cite{LARGE_PAP}.

\begin{figure}
\begin{center}
\vspace{-0.2cm}
\includegraphics{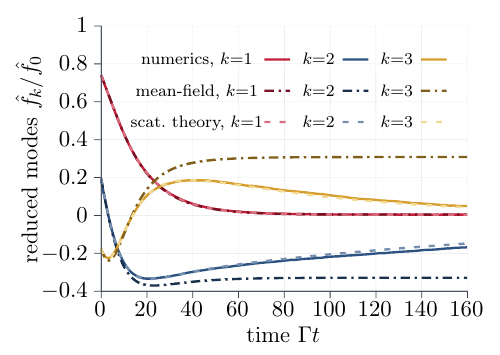}
\vspace{-0.5cm}
\caption{Reduced angular modes $\hat{f}_k/\hat{f}_0$ versus dimensionless time
$\tilde{t}=t\,|\Gamma|$, from
agent-based simulations (solid), mean-field (dotted),
and scattering theory (dashed) for $S=0.05$, $M=0.1$,
$N=493$, $\Gamma=-0.2$,
time step $\dd t=0.025$, $v_0=4$, $R=1$, $L=124$,
$\eta=75^\circ$. Mean-field theory predicts a decay into an unphysially ordered state for the anti-aligning model.
}
\label{FIG_RELAX1}
\vspace{-0.9cm}
\end{center}
\end{figure}

We consider $N$ identical SPPs in two dimensions moving at constant speed $v_0$,
\begin{equation}
\label{POS_ANGLE_EQ}
\frac{\dd\vec{r}_i}{\dd t}=v_0\,\uvec{n}(\theta_i)\,,\quad
\frac{\dd\theta_i}{\dd t}=\Gamma\!\sum_{j\in\Omega_i}
\sin(\theta_j-\theta_i) + \zeta_i(t)
\end{equation}
with weak white Gaussian noise, $\langle \zeta_i(t)\rangle {=}0$, $\langle \zeta_i(t) \zeta_j(t')\rangle {=}2D_r\delta_{ij}\delta(t{-}t')$. Particle $i$ sits at $\vec{r}_i$ and flies along $\uvec{n}(\theta_i)=(\cos{\theta_i},\sin{\theta_i})$; the set $\Omega_i$ contains all particles within the interaction range $R$ of it. For $\Gamma>0$, the coupling is aligning and produces global polar order, \emph{flocking}, at sufficiently strong coupling or weak noise; for $\Gamma<0$, it is anti-aligning, favouring ``social distancing''~\cite{borra2021,nourhani2021,alert2024,escaff2024,*escaff2024b,boltz2025,bektas2025,ketzetzi2025,lardet2026,schadschneider}. Under anti-alignment, two interacting particles separate after a time of order $\tau_{\text{coll}}= R/v_0$, whereas nearly parallel aligning particles can remain bound arbitrarily long. The collision picture is therefore naturally suited to anti-alignment; alignment requires a separate regularisation, discussed below. We thus assume $\Gamma<0$ unless stated otherwise, so that $R$ and $\tau_{\text{coll}}$ set the natural scales. The dimensionless parameters are the partner number $M=\pi R^2 \rho_0$, with number density $\rho_0=N/L^2$ in a box of size $L\times L$, and the coupling strength $S=\Gamma R/v_0$.

Solving \cref{POS_ANGLE_EQ} formally for the positions,
$\vec{r}_i(t)=\vec{r}_i(0){+}v_0\int_0^t \dd\tilde{t}\,
\uvec{n}(\theta_i(\tilde{t}))$,
yields an evolution equation for the angles alone,
$\dot{\theta}_i=\Gamma\,\sum_{j=1}^N a_{ij}
\sin(\theta_j-\theta_i)$, with adjacency matrix
$a_{ij}\equiv \Theta(R-\lvert\vec{r}_i(t)-\vec{r}_j(t)\rvert)$.
Through the positions, $a_{ij}$ is a functional of the past angles.
Thus, this is an XY-type rotator network with history-dependent topology,
for which, to our knowledge, no solutions exist.
Collecting the microscopic state in
$\vec{Z}\equiv(\vec{r}_1,\theta_1,\ldots,\vec{r}_N,\theta_N)\equiv(\ph{1},\ldots,\ph{N})$,
where $\ph{j}$ abbreviates the phase of particle $j$, the probability density
$P_N(\vec{Z},t)$ obeys the Fokker--Planck equation (Liouville eq. for $D_r=0$),
\begin{align}
\begin{split}
\partial_t P_N =&-\sum_{i=1}^N\Big\{ \left[v_0\,(\uvec{n}_i\cdot\vec{\nabla}_i) - D_r \partial_{\theta_i}^2 \right]\, P_N \\
&+\partial_{\theta_i}\,\sum_{j=1}^N \Big[\Gamma\, a_{ij}\, \sin(\theta_j-\theta_i)
\Big]P_N \Big\} \end{split} \,.
\label{N_FOKK}
\end{align}
Marginalising over $N-k$ phases converts this into a BBGKY hierarchy~\cite{balescu_75}
for the $k$-body distributions,
$f_k(\ph{1},\ldots,\ph{k},t) \equiv \frac{N!}{(N-k)!}
\int P_N \,\dd\ph{(k+1)}\cdots \dd\ph{N}$.
Integrating over $N-1$ and $N-2$ phases, respectively, gives
\begin{align}
\label{FIRST_MEM}
\begin{split}
&\partial_t f_1=-v_0\uvec{n}(\theta)\cdot\vec{\nabla} f_1 + D_r \partial_\theta^2 f_1 \\
&{-}\Gamma
\partial_{\theta}\!\!\iint\! \dd\theta_2 \dd\vec{r}_2\,
a(\lvert\vec{r}_2{-}\vec{r}\rvert)\,\sin(\theta_2{-}\theta)\,f_2(\vec{r}{,}\theta{,}\vec{r}_2{,}\theta_2{,}t)\end{split}\\
\begin{split} &\partial_t f_2 ={-}v_0\left[\uvec{n}(\theta)\cdot\partial_{\vec{r}}{+}\uvec{n}(\beta)\cdot\partial_{\vec{z}} \right] f_2 + D_r (\partial_\theta^2{+}\partial_\beta^2) f_2 \\
&{-}\Gamma\,a(\lvert\vec{r}-\vec{z}\rvert) \, \big[\big(\partial_{\theta}{-}\partial_{\beta}\big)\,\sin(\beta  -\theta)\,f_2\big]
{-}\Gamma\!\int\!\dd\ph{3} \,F[f_3] \end{split}
\label{SECOND_MEM}
\end{align}
with $F\equiv [a(\lvert\vec{r}_3-\vec{r}\rvert) \partial_{\theta}\, \sin(\theta_3  -\theta)
+a(\lvert\vec{r}_3-\vec{z}\rvert) \partial_{\beta}\, \sin(\theta_3  -\beta)]
f_3(\vec{r},\theta,\vec{z},\beta,\vec{r}_3,\theta_3,t)$.

The simplest closure
factorises the probability density, $f_k = \prod_{j=1}^k f_1(\ph{j})$, the mean-field
approximation of \emph{molecular chaos}, leaving a closed equation for the one-particle
density $f(\vec{r},\theta,t)\equiv f_1(\ph{1})$. This Ansatz must fail, however.
Consider two particles at the same location $\vec{r}$ for $M\ll 1$ in the deterministic
limit $D_r \to 0$. For them to be this close, they must have interacted for some time,
during which anti-alignment turned their directions apart, so equal directions are
impossible, $f_2(\vec{r},\theta,\vec{r},\theta,t)=0$. As $f_1$ is non-zero everywhere,
factorisation cannot hold, $0=f_2\neq f_1^2>0$. The two-particle phase
space thus contains ``forbidden'' zones, among them the points
$\vec{r}_1\approx\vec{r}_2$, $\theta_1\approx\theta_2$;
see~\cite{LARGE_PAP} for a detailed evaluation.

We now quantify this failure. In terms of angular Fourier modes,
$\hat{f}_n(\vec{r},t) {=}  \int_0^{2\pi} \mathrm{e}^{-\ii n\theta}f(\vec{r},\theta,t)\,\dd\theta/2\pi$,
\cref{FIRST_MEM} becomes a mode hierarchy with collision integral $J^{\text{MF}}_n$,
\begin{equation}
\label{MF_KINETIC1}
\frac{\DD\hat{f}_n}{\DD t}=J^{\text{MF}}_n[\hat{f},\bar{\hat{f}}]\equiv
A n \pi \Gamma \, \left\{
\hat{f}_{n-1}\bar{\hat{f}}_1-\hat{f}_{n+1}\bar{\hat{f}}_{-1}
\right\} \,,
\end{equation}
where $A=\pi R^2$ is the area of the collision circle and
$\bar{\hat{f}}_k(\vec{r})\equiv \int a(|\vec{r}_2-\vec{r}|)\,\hat{f}_k(\vec{r}_2)\,\dd\vec{r}_2/A$
is the average of $\hat{f}_k$ over it. The material derivative
$\DD \hat{f}_n/\DD t\equiv\partial_t\hat{f}_n+\frac{v_0}{2}[\nabla^*\hat{f}_{n-1}+\nabla \hat{f}_{n+1}]$
contains the complex gradient $\nabla=\partial_x+\ii \partial_y$ and its conjugate.
The modes $\hat{f}_0=\rho/2\pi$ and $\hat{f}_1$ give the number density and the polar
order, the real and imaginary parts of the latter being the $x$- and $y$-components of
the momentum density. For a homogeneous, unpolarised state, $\hat{f}_1=0$,
\cref{MF_KINETIC1} predicts $\partial_t \hat{f}_n=0$ for $|n|\ge 2$: higher modes never
relax. This is not merely counterintuitive; it contradicts agent-based simulations of
\cref{POS_ANGLE_EQ}, in which $N$ particles are placed uniformly on a torus with
directions drawn from $[-\eta,\eta]$, so that initially
$\hat{f}_n=(\rho_0/2\pi)\,\operatorname{sinc}(n\,\eta)$, and the modes are measured as
$\hat{f}_n(t)=\left\langle \sum_j \exp(-\ii n\theta_j(t))/2 \pi L^2\right\rangle$,
averaged over an ensemble of runs. \Cref{FIG_RELAX1} shows all modes relaxing to zero,
whereas mean-field theory freezes those with $|n|>1$ at non-zero values.

For finite $D_r$, a term $-n^2 D_r \hat{f}_n$ is added to \cref{MF_KINETIC1}; absent
global order, this diffusive decay dominates at long times and obscures the failure.
A second, independent motivation for going beyond mean-field is that its
collision integral can vanish altogether: where the interaction depends on the angle
between orientation and displacement vector, an additional rotational symmetry makes the
mean field vanish identically~\cite{negi2022,alert2024,negi2024,huang2025}.
We therefore replace molecular chaos by \emph{one-sided molecular
chaos}~\cite{kreuzer_81,waldmann_58}, neglecting correlations~\cite{kuersten2020,murphy2024} immediately before a
collision~\cite{soto2024} but determining them explicitly during it.

To obtain a theory beyond mean-field, the second hierarchy equation, \cref{SECOND_MEM}, is considered where
the three-body function 
$f_3$ contributes only if its spatial coordinates are not further apart than $2R$
from each other. In the limit of small densities, $M\rightarrow 0$, the probability of three-particle collisions is negligible, 
allowing us to set $f_3=0$. 
This \emph{binary-collision approximation} reduces the BBGKY-hierarchy to  
two equations. Next, we further reduce the theoretical description to a single 
kinetic equation. 
To evaluate the first hierarchy equation, 
it suffices to know the two-body probability density $f_2$ inside
the collision circle, i.e., for $|\vec{r}_2-\vec{r}|\leq R$.
With this in mind, and neglecting noise during individual collisions, which is valid for $D_r \tau_{\text{coll}} \ll 1$, the second hierarchy equation reduces to the Liouville equation of two interacting bodies ($a_{12}\equiv 1$), and can be solved by the method of characteristics using the actual particle trajectories.
The microscopic equations \eqref{POS_ANGLE_EQ} 
are solved exactly for two particles, yielding the time dependence of the difference angle 
$\Delta(t)\equiv \theta_2(t)-\theta_1(t)= 2\,\operatorname{atan}\{\tan[\Delta(t_0)/2]\,\exp[2\Gamma(t_0-t)]\}$, whereas 
$\theta_1+\theta_2$ remains constant in time. 
The differential equation 
$\dd f_2/\dd t=-\Lambda_2\, f_2$
describes the evolution of $f_2$
along the characteristics with
the phase space compression factor \cite{evans_08} of 
the two-particle system,
$\Lambda_2=-2\Gamma \,\cos\Delta$. This is an explicit reminder of the Non-Hamiltonian nature of the dynamics: the probability density is not invariant along particle trajectories. Solving this equation gives
\begin{align}
\begin{split}
 &f_2(\vec{r}(t){,}\theta(t){,}\vec{r}_2(t){,}\theta_2(t){,}t)=
\mathrm{e}^{2\Gamma\!\!\int_{t_0}^t\!\! \dd\tilde{t} \cos\{\theta_2(\tilde{t})-\theta(\tilde{t})\}} \\
 & \qquad\qquad\qquad\times f_2(\vec{r}(t_0),\theta(t_0),\vec{r}_2(t_0),\theta_2(t_0),t_0)
\end{split}\,.\label{SUPER_POS}
\end{align}
Here, the notation emphasises that we follow the phases along actual trajectories from time $t_0$ to the time of interest $t$. Identifying $t_0$ as the time two particles start to interact, one-sided molecular chaos 
is employed by factorising the initial condition, 
$f_2(\ph{1},\ph{2},t_0)\approx f_1(\ph{1},t_0)\,f_1(\ph{2},t_0)$.
For two particles currently at $\vec{r}_1{=}\vec{r},\theta_1{=}\theta,\vec{r}_2,\theta_2$
with $|\vec{r}_2-\vec{r}|\le R$, tracing the trajectories back to
$|\vec{r}_2-\vec{r}|=R$ determines $t_0$, the collision time
$t_{\text{coll}}=t-t_0$, and the entrance phases exactly~\cite{LARGE_PAP}. Inserting the
resulting $f_2$ into \cref{FIRST_MEM} effects a \emph{non-local} closure of the kinetic equation.
In the dilute limit, $M\to 0$, the average duration of a collision $\langle t_{\text{coll}}\rangle \sim \tau_{\text{coll}}$ is 
much smaller than the time between collisions, and the radius of interaction, $R$, is much smaller than the mean free path.
Aiming for a coarse-grained description on time  and length scales above the microscopic scales $\tau_{\text{coll}}$ and $R$, respectively, we approximate $t\approx t_0$ and
$\vec{r}(t)\approx \vec{r}_2(t), \vec{r}(t_0), \vec{r}_2(t_0)$  in \cref{SUPER_POS}.

With these simplifications 
we evaluate
the first hierarchy equation perturbatively for small coupling $S$. This is not a fundamental restriction, but we limit the presentation here to analytically tractable integrals. The theory is asymptotic in density but not in coupling and can be evaluated at
arbitrary $S$~\cite{mihatschlarge}, unlike Landau-eq.
methods~\cite{patelli,boltz2024,kuersten2024}, which are perturbative in a combination of both. In angular Fourier
space, $J^{\text{MF}}_n$ is recovered at first order in $S$, where the forbidden zone
does not contribute. At $\mathcal{O}(S^2)$ it does, adding 
$J_m^{(2)}[\hat{f},\bar{\hat{f}}]=R v_0\,S^2\,\sum_{n=-\infty}^{\infty}\hat{f}_{m-n}\bar{\hat{f}}_{n}\,
g_{mn}$  
with coupling matrix
\begin{equation}
\label{G_MN_DEF}
g_{mn}=\frac{8}{3} m\left[ \frac{\frac{1}{2}m+n}{n^2-\frac{1}{4}}+\frac{\frac{3}{2}m-n}{n^2-\frac{9}{4}}\right] \,.
\end{equation}
Thus, the binary-collision approximation results in 
the following mode hierarchy, the Boltzmann equation,
\begin{align}
\label{MF_KINETICFULL}
\begin{split}
\frac{\DD \hat{f}_n}{\DD t} =&\, J^{\text{MF}}_n[\hat{f},\bar{\hat{f}}]
+J^{(2)}_n[\hat{f},\bar{\hat{f}}]-n^2 D_r \hat{f}_n\,, 
\end{split}
\end{align}
where we included the diffusive effect of the angular noise and are missing terms $\mathcal{O}(S^3,M^2, (D_r\tau_{\text{coll}})^2)$. As a result of our coarse-graining, contributions from the noise during the collisions are omitted. A perturbative operator expansion can be considered,~\cite{boltzdyson} 
but we are interested in the case of weak noise here, so that the leading order suffices. 
As discussed before, the regular mean-field closure, i.e., the omission of  
$J^{(2)}$
leads to the unphysical result that the higher modes do not relax to zero.
We show in \cref{FIG_RELAX1} that including this term leads to excellent quantitative agreement with the agent-based results over the entire range of the measurement time. 

\begin{figure}
\centering
\includegraphics[width=\linewidth]{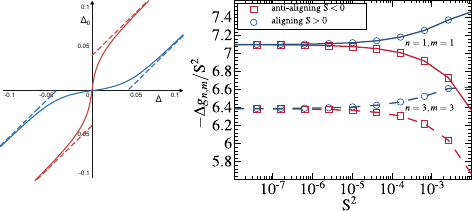}
\vspace*{-0.6cm}
\caption{Left: Back-tracing map $\Delta_0(\Delta)$ with (solid) and without speed disorder regularizations ($\omega^2/v^2=0.03$); colours same as right panel. Right: Numerically determined deviation of the coupling elements from mean-field theory with speed regularization ($\omega/v=0.0256$) as a function of the quadratic coupling $S^2$. See text for discussion and ref.~\citenum{LARGE_PAP} for details.}
\label{fig:disorder}
\vspace{-.5cm}
\end{figure}

The discussion so far has focussed on anti-alignment, $\Gamma<0$. The mean-field
term is trivially the same for either sign of~$\Gamma$. For higher orders, barring
phase transitions (in the disordered phase, i.e., for sufficiently weak
coupling at finite $D_r$), physical quantities must be analytic in~$S$. Reading
\cref{MF_KINETICFULL} as an expansion in $S^2$, the same expression should hold for
both signs. Two features of the anti-aligning case, however, do not carry over: first,
there is no forbidden zone for aligning particles; second, the collision times diverge
for vanishing angle difference within the deterministic binary collision approximation and are only broken up by external noise or multi-particle events. Na\"{i}vely adapting the integration domains to account
for the absent forbidden zone produces an additional $\mathcal{O}(S^2)$ term in the
collision integral. This contradicts the analyticity argument. Moreover, the collision
integral as derived conserves total probability; the spurious contribution would violate
this~\cite{LARGE_PAP}. Such secular terms that break the symmetries of the full theory
are common in operator expansions and must be treated with care.

The physical origin is clear: regions of phase space with diverging interaction times
contribute, undermining the binary collision picture. One could excise these by hand,
either by imposing analyticity or by appealing to finite angular noise, $D_r>0$, which
eventually dissolves any bound pair. Neither route is fully satisfying. Instead, we
introduce a \emph{physical regularisation} based on \emph{quenched speed noise}. If two
colliding particles have different speeds, $v_1=v_0-\omega/2$ and $v_2=v_0+\omega/2$,
their relative velocity
$v_{\text{rel}}=\sqrt{4 v_0^2 \sin^2\!\frac{\Delta}{2}+\omega^2\cos^2\!\frac{\Delta}{2}}$
is nonzero for any $\omega>0$, guaranteeing a
finite collision time. For the binary calculation only a single pair is needed, but the
physical picture is a polydisperse variant of \cref{POS_ANGLE_EQ} with $v_0\to
v_i=v_0+\omega\nu_i$, where the $\nu_i$ are quenched random numbers of zero mean drawn,
e.g., from a uniform distribution on $[-\tfrac{1}{2},\tfrac{1}{2}]$. The full
calculation is in the companion paper~\cite{LARGE_PAP}; in the regularised limit
$\omega\to 0^+$, \cref{MF_KINETICFULL} holds for either sign of $\Gamma$. Any
finite~$\omega$ eliminates the non-analytic behaviour of the back-tracing map
$\Delta_0(\Delta)$: the forbidden zone for $\Gamma<0$ and its time-reversed
counterpart, a zone of unreachable initial configurations for $\Gamma>0$, cf.\
\cref{fig:disorder} (left). That these two zones must be related follows from the
symmetry $\Gamma\to-\Gamma$, $t\to-t$. Direct numerical evaluation of the collision
integral confirms that the asymptotic corrections to mean-field agree for both signs at
small $S^2$, see \cref{fig:disorder} (right).

As an application, we derive the effective Langevin equation for the angular dynamics
at $\Gamma<0$, including the effective noise which is a quantity of interest often treated only
by semi-heuristical arguments~\cite{spera2024} which have been found~\cite{steinhofel2026} to miss important physics. Following
Boltzmann--Lorentz theory~\cite{hauge_70,brey_99,garzo_03}, a single tagged particle
($i=1$) with density $h(\vec{r},\theta)\equiv\langle\delta(\vec{r}-\vec{r}_1)\,
\delta(\theta-\theta_1)\rangle$ is singled out from the $N-1$ background particles
whose density obeys \cref{MF_KINETICFULL}. The tagged density fulfils the linear
 equation
$\DD\hat{h}_n/\DD t=J^{\text{MF}}_n[\hat{h},\bar{\hat{f}}]
+J^{(2)}_n[\hat{h},\bar{\hat{f}}]-n^2 D_r \hat{h}_n$.
For a homogeneous, disordered background, $\hat{f}_n=\delta_{n,0}\,\rho_0/2\pi$,
the velocity autocorrelation function (VAF)
\begin{equation}
\label{AUTO_CN}
C(\tau)\equiv\langle \vec{v}_i(t+\tau)\cdot \vec{v}_i(t)\rangle
=v_0^2\langle \cos(\Delta \theta)\rangle
\end{equation}
decays exponentially (see ref.~\citenum{LARGE_PAP} and appendix), $C(t)=v_0^2\exp(-t/\tau_C)$, with
$\tau_C^{-1}=(32R\Gamma^2M)/(9\pi^2v_0)+D_r$ and self-diffusion coefficient
$D=\tau_C v_0^2/2$. Mean-field theory alone cannot capture self-mixing and gives only
the $D_r$ contribution. In the deterministic limit, it therefore predicts a diverging self-diffusion coefficient.

\begin{figure}
\begin{center}
\vspace{-0.cm}
\includegraphics[width=.68\linewidth,angle=0]{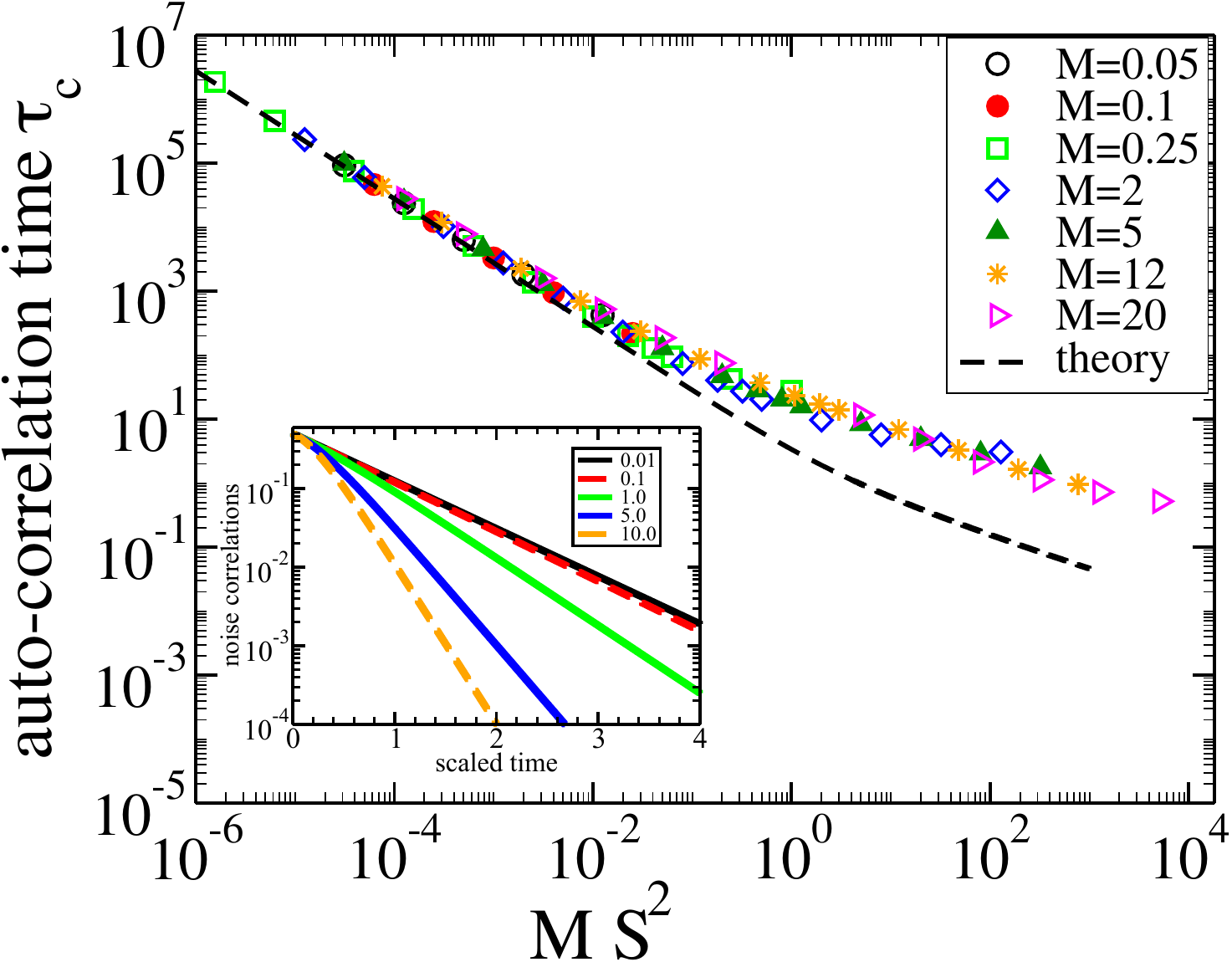}
\vspace{-0.3cm}
\caption{
Correlation time $\tau_C\,v_0/R$ versus $M\,S^2$ from agent-based simulations (symbols)
compared to theory, $\tau_C=2/\sigma^2$, using \cref{SIGMA_SELF_CONS}.
Inset: noise correlations $\langle \xi(t)\xi(0)\rangle/\gamma$ versus time $t\,v_0/R$, from \cref{NOISE_EXACT}.
\vspace{-0.9cm}
}
\label{FIG2_AUTOCORR}
\end{center}
\end{figure}

Writing the angular dynamics as a Langevin equation $\dot\theta=\xi$ with Gaussian noise
and comparing the resulting VAF to the kinetic prediction determines the effective
white-noise strength in the dilute limit,
\begin{equation}
\label{RELAT_ANG_SIGMA1A}
\sigma^2=\frac{64 R\, \Gamma^2\, M}{9 \pi^2 v_0}+2D_r\;\;\;\;
\text{for}\;\;M,\,D_r\tau_{\text{coll}}\ll 1 \,.
\end{equation}
To extend to arbitrary $M$, we model the adjacency-matrix elements $a_{ij}$ as
independent \emph{random telegraph processes}~\cite{vanKampen-book} with
\textsc{off}-rate $w_{\text{off}}=B\,v_0/R$ (the \textsc{on}-rate is negligible for
$N\gg 1$) and telegraph correlation
$\langle a_{ij}(t{+}\tau)\,a_{ik}(t)\rangle=\delta_{jk}\,M\,
\mathrm{e}^{-w_{\text{off}}|\tau|}/N$.
Neglecting correlations between $a_{ij}$ and the angles~\cite{LARGE_PAP}, the
mean-square angular displacement satisfies
$x(\tau)=
\frac{M\Gamma^2}{2}\!
\int_0^{\tau}\!\! \dd\tilde{t}
\int_0^{\tau}\!\! \dd t'\,
\mathrm{e}^{-w_{\text{off}}|\tilde{t}-t'|-x(\tilde{t}-t') }\,,
$ 
which, upon differentiating twice, yields $\ddot{x}=\gamma\,
\mathrm{e}^{-x-w_{\text{off}}|\tau|}$ with $\gamma\equiv M\Gamma^2$, solved exactly by
\begin{equation}
\label{EXACT_SOL}
x=2\,\ln\!\left\{
(c^2+\mathrm{e}^{-2\lambda|\tau|})(
2b\,c\,\mathrm{e}^{-\lambda|\tau|})^{-1}\right\}
-w_{\text{off}}|\tau|\,,
\end{equation}
with $\epsilon\equiv\gamma/w_{\text{off}}^2$, $b^2=1+1/(2\epsilon)$,
$c=b+\sqrt{b^2-1}$, $\lambda=w_{\text{off}}\,b\sqrt{\epsilon/2}$.
The noise correlations are
\begin{equation}
\label{NOISE_EXACT}
\langle\xi(t)\xi(\tilde{t})\rangle
=\frac{\gamma}{2}\!\left[
2bc\,\mathrm{e}^{-\lambda|t-\tilde{t}|}(
c^2+\mathrm{e}^{-2\lambda|t-\tilde{t}|})^{-1}
\right]^2\,,
\end{equation}
which become exponential for $|\tau|\gtrsim\tau_{\text{coll}}$ (\cref{FIG2_AUTOCORR},
inset). Coarse-graining on this scale gives effective white noise with strength
\begin{equation}
\label{SIGMA_SELF_CONS}
\sigma^2=\lim_{\tau\to\infty}\dot{x}(\tau)
=w_{\text{off}}\!\left[\sqrt{1+2\epsilon}-1\right]\,.
\end{equation}
The noise is controlled by $\epsilon\sim MS^2$; matching to
\cref{RELAT_ANG_SIGMA1A} at small $\epsilon$ fixes $B=9\pi^2/64$.
\Cref{FIG2_AUTOCORR} shows that the prediction agrees well with
simulations at small $MS^2$. At larger $M$ or $S$ the theory underestimates $\tau_C$,
likely because the telegraph approximation neglects multi-particle correlations inside
the collision circle.

We have constructed an asymptotically exact kinetic theory for dilute aligning matter by closing the BBGKY hierarchy non-locally with the exactly solved two-body problem. The resulting Boltzmann equation reproduces the relaxation of the angular mode hierarchy quantitatively, where mean-field theory fails qualitatively. Because the collision integral follows from the equations of motion, its form and coefficients are fixed by the microscopic model and specific to it rather than chosen. For aligning couplings, where collision times diverge, a regularisation by quenched speed disorder yields the same collision integral, as required by analyticity. Divergences of this kind are familiar from classical kinetic theory, where the Landau equation requires small-angle cutoffs and Balescu--Lenard resolves the large-scale divergence through dynamical screening~\cite{balescu_75}. Here, the divergence is in the collision duration and it is removed by an arbitrarily small speed polydispersity, a regularisation that should apply whenever bound pair states threaten the binary-collision picture. A self-consistent mapping of the interaction network to a random-telegraph process extends the description to arbitrary density and gives an effective one-particle Langevin equation with analytic expressions
for its noise correlations and the self-diffusion coefficient. Furthermore, the Boltzmann level allows for more physical  Toner-Tu type~\cite{toner1998,*toner} hydrodynamic equations, see appendix.

Beyond SPPs, \cref{POS_ANGLE_EQ} is a network of rotators whose topology is generated by the dynamics itself. Coupled rotators are studied as models of neural and cardiac tissue~\cite{pikovsky_01,buzsaki_06}, where the asynchronous state is typical~\cite{poulet_08,harris_11,vrees_96} and units behave quasi-stochastically
because they are driven by many other units. For permanently but
randomly coupled rotators, effective Langevin equations have been established within
stochastic mean-field approximations~\cite{sompolinsky_88,sompolinsky_82,stiller_98,schuecker_16,vanmeegen_18,ranft_22a,ranft_22b,kati2024}. Here the coupling matrix is not quenched but a dynamic functional of the angle history, and the resulting network noise is obtained analytically.
 
The construction requires only that binary collisions dominate; the microscopic interaction is otherwise unconstrained. Non-reciprocal~\cite{mihatsch2025}, nematic and multi-species~\cite{kuersten_23} couplings as well as chiral~\cite{liebchen2022} or self-aligning dynamics~\cite{baconnier2025,wagner} are therefore within reach. The dilute strong-coupling regime to which this method is particularly suited is relevant for cohesion~\cite{shea2025} or turn-away-type models~\cite{alert2024} as well as vision-cone interactions~\cite{negi2022,negi2024,huang2025}. These types of models that couple as a function of the relative displacement are typically non-reciprocal and often feature vanishing mean-field contributions. On a broader scope, the alignment interaction studied here is just one realization of interaction in internal variables and the method applies also to other forms of such interactions, for example learning in smart active matter~\cite{jung2025} or opinion formation~\cite{wolfram2025}. Another extension that would require higher dimensionality is to include inertial effects~\cite{tevrugt2026b}.  
 The present approach is not perturbative in the coupling, and evaluating the collision integral at finite $S$ is a natural next step.

\pagebreak \onecolumngrid
\appendix
\section{End Matter: Technical Derivations}
\subsection{Velocity autocorrelation}

In this appendix, we show that the derived Boltzmann equation used within a Boltzmann--Lorentz framework can be used to derive the velocity autocorrelation function and, thus, the self-diffusion coefficient. We start by inspecting \cref{AUTO_CN} in more detail
\begin{align}
&C(\tau)\equiv\langle \vec{v}_i(t+\tau)\cdot \vec{v}_i(t)\rangle\notag\\
&= v_0^2 \int\dd\vec{r}\dd\theta\dd\vec{r}_0\dd\theta_0 p(\vec{r},\theta,t+\tau | \vec{r}_0,\theta_0,t_0) \cos(\theta-\theta_0)=v_0^2 \int\dd\vec{r}\dd\theta p(\vec{r},\theta,t+\tau | 0,0,t_0) \cos(\theta) \text{.}
\end{align}
In the last step, we used the symmetry of the propagator and renamed variables. An important observation is that the single-particle propagator $p(\vec{r},\theta,t+\tau | 0,0,t_0)$ is given by the solution $h$ to the kinetic Boltzmann-Lorentz equation for the single-particle function of a tagged particle with appropriate initial conditions. In particular, we use (and already have used) uniformity in space and angles. If we choose the reference frame for the angle such that the Fourier components $\hat{h}_n$ of $h$ are real (such that $\hat{h}_n=(\hat{h}_{-n}^*)^*=\hat{h}_{-n}$), we directly get
\begin{equation}
    C(\tau)=\hat{h}_1(\tau)
\end{equation}
and the Boltzmann--Lorentz equation for $\hat{h}_1$ in a disordered background, i.e., $\hat{f}_n = \overline{\hat{f}}_n = \delta_{n,0} \frac{\rho}{2\pi}$, is
\begin{equation}
\partial_t \hat{h}_1 = (\frac{\rho_0}{2\pi} g_{1,0} -D_r)\hat{h}_1
\end{equation} 
and the result from the main text follows.

\subsection{Toner-Tu type hydrodynamics}

The purpose of this section is to show how to derive hydrodynamic equations of Toner-Tu type~\cite{toner1998,*toner} from the Boltzmann equation. To this end, we make some assumptions that allow for a short presentation here, while a more careful analysis will be presented elsewhere. First, we assume that the density $\rho$ and the velocity field $\vec v$ are the relevant modes and there is a meaningful closure of the hydrodynamics in these. Second, we neglect variability on the coarse-graining length-scale $R$ meaning we do not distinguish between $\hat{f}$ and its locally averaged counterpart $\overline{\hat{f}}$. This will lead to an omission of higher order gradient terms that could be derived in principle. Third, we assume that all the neglected modes that do not contribute to the remaining physical fields $\rho$ and $\vec v$ are small homogeneous constants and can be determined via linearisation. Specifically, these are modes $\hat{f}_n$ with $\lvert n \rvert\geq 2$ as $n=0$ corresponds to the density and we can build the velocity field taking also $n=\pm 1$ into account. Noting that the mean-field collision integral has the same structure as the genuine Boltzmann terms, we can write mode equations as
\begin{equation}
\partial_t \hat{f}_n + \frac{v_0}{2} (\nabla^* \hat{f}_{n-1} + \nabla \hat{f}_{n+1}) = K_{nm} \hat{f}_{n-m} \hat{f}_m - n^2 D_r \hat{f}_{n}
\end{equation}
with coupling elements $K_{nm}$ that include both the mean-field contributions and the Boltzmann coupling matrix $g_{nm}$. Restricting ourselves to $n\geq 2$ as the negative case follows from the general relation $\hat{f}_{-n}=\hat{f}_n^*$, we see that our assumptions allow us to set the left-hand side to zero and drop most of the terms of the right-hand side. What remains is a recurrence relation for $n \geq 2$
\begin{equation}
\hat{f}_n = \alpha_n \hat{f}_0 \hat{f}_n + \beta_n \hat{f}_{-1} \hat{f}_{n+1} + \gamma_n \hat{f}_1 \hat{f}_{n-1}
\end{equation}
with coefficients $\alpha_n = (n^2 D_r)^{-1} ( K_{n,0}+ K_{n,n})$, $\beta_n = (n^2 D_r)^{-1} ( K_{n,n+1}+ K_{n,-1})$ and $\gamma_n = (n^2 D_r)^{-1} ( K_{n,n-1}+ K_{n,1})$ with the understanding that only one coupling element is to be taken in cases where the two index combinations in one coefficient coincide. We have separated the slow modes explicitly to make the dependence on these obvious. Introducing ratios $R_n = \frac{\hat{f}_n}{\hat{f}_{n-1}}$, we find
\begin{equation}
    R_n = \frac{\gamma_n \hat{f}_1}{1-\alpha_n \hat{f}_0-\beta_n \hat{f}_{-1} R_{n+1}} \text{.}
\end{equation}
This allows for a representation of the $R_n$ as continued fractions, but another way allows for more direct insight. We write $R_n=\gamma_n\hat{f}_1/C_n$ to find a recurrence relation for the divisors
\begin{equation}
C_n = 1-\alpha_n \hat{f}_0-\beta_n \hat{f}_{-1}\hat{f}_1 \gamma_{n+1} / C_{n+1}
\end{equation}
Noting that $\hat{f}_{n\geq 2} = \hat{f}_1 \prod_{k=2}^n R_k$ and the structure of the divisors $C_n$, we see that
 \begin{equation}
 \hat{f}_{n\geq 2} = \hat{f_1}^{n} q_n(\hat{f}_0,\hat{f}_{1}\hat{f}_{-1})
 \end{equation} 
 with $q_n$ an analytic function in both $f_0$ and the combination $\hat{f}_{1}\hat{f}_{-1}=\lvert\hat{f}_1\rvert^2$.  This result gives some credence to the assumptions: if we are interested in the emergence of polar transport from a disordered state, the velocity modes $\hat{f}_{\pm 1}$ are going to be small, and these higher modes are suppressed by additional powers of this small quantity. Restricting the analysis to second order in velocity, we see that we have to only determine $\hat{f}_2$ (and then by analogy $\hat{f}_{-2}$). We find
 \begin{equation}
 \hat{f}_2 = \gamma_2 \hat{f}_1^2 \sum_{k=0}^\infty (\alpha_2\hat{f}_0)^k 
 \end{equation}
where we can safely assume the geometric series to converge as we are interested in the dilute limit. This scaling is in line with the results from Chapman-Enskog style expansions~\cite{patelli,LARGE_PAP}. With this, we have a closed set of equations
\begin{align}
\partial_t \hat{f}_{-1} + \frac{v_0}{2} (\nabla^* \hat{f}_{-2} + \nabla \hat{f}_{0}) &= (K_{-1,-1} + K_{-1,0} )\hat{f}_{-1} \hat{f}_{0} + (K_{-1,1}+K_{-1,-2}) \hat{f}_{-2} \hat{f}_{1} -  D_r \hat{f}_{-1} \\
\partial_t \hat{f}_0 + \frac{v_0}{2} (\nabla^* \hat{f}_{-1} + \nabla \hat{f}_{1}) &= 0 \label{eq:appcont}\\
\partial_t \hat{f}_1 + \frac{v_0}{2} (\nabla^* \hat{f}_{0\phantom{-}} + \nabla \hat{f}_{2}) &= (K_{1,1\phantom{-}\phantom{-}} + K_{1,0\phantom{-}} )\hat{f}_{1\phantom{-}} \hat{f}_{0} + (K_{1,-1}+K_{1,2\phantom{-}\phantom{-}}) \hat{f}_{2\phantom{-}} \hat{f}_{1} -  D_r \hat{f}_{1}
\end{align}
Since we decided to truncate at quadratic order in the velocity, we can drop all $\hat{f}_{\pm 2}$ terms on the right-hand side. We note that $\rho=\hat{f}_0 2\pi$ and $\rho \vec v = v_0 \pi (\hat f_{1} + \hat f_{-1}, \ii(\hat f_{1} - \hat f_{-1}) )^\mathrm{T}$. Thus, \cref{eq:appcont} is the continuity equation. For the $\vec v$ dynamics we find using $\alpha=\rho (K_{1,1} + K_{1,0} )/(2\pi)-D_r=\rho (K_{-1,-1} + K_{-1,0} )/(2\pi)-D_r$ and $\gamma=\gamma_2/(2\pi)=\gamma_{-2}/(2\pi)$ as shorthands and cutting $\hat{f}_2$ at zeroth order in density
\begin{align}
\partial_t \vec v &= \alpha \vec v + \vec v(\nabla\cdot\vec v) + \gamma \rho  \left[ \frac{1}{2}\nabla v^2 - \vec v (\nabla\cdot\vec v) - (\vec v \cdot \nabla) \vec v\right] + \vec{v}(\vec v\cdot \frac{\nabla\rho}{\rho})-\frac{v_0^2}{2} \frac{\nabla \rho}{\rho} + \gamma  \left[ - 2 \vec v(\vec v\cdot \nabla \rho) + v^2 \nabla \rho\right]
\end{align}
or in more standard form using $v^2\nabla\rho=\frac{1}{\rho} \nabla(\frac{\rho^2v^2}{2})-\frac{\rho}{2}\nabla v^2$ and $\vec{v}\cdot(\vec{v}\cdot \nabla\rho)=\frac{1}{\rho} \vec v \cdot (\vec{v} \cdot \nabla \frac{\rho^2}{2})$
\begin{equation}
\begin{aligned}
\partial_t \vec v &+ \lambda_1 (\vec v \cdot \nabla) \vec v + \lambda_2(\nabla\cdot\vec v)\vec v + \lambda_3 \nabla v^2 = U  \vec v - \frac{1}{\rho} \nabla P  - \frac{1}{\rho}\vec{v}(\vec v\cdot\nabla P_2) \qquad \text{with}\\
&\begin{alignedat}{4}
\lambda_1 &= \gamma\rho, &\qquad \lambda_2 &= -1+\gamma\rho, &\qquad \lambda_3 &= 0,\\
 U &= \alpha, &\qquad P &=\rho v_0^2/2-\gamma\rho^2v^2/2, &\qquad P_2 &= -\rho+\gamma\rho^2
\end{alignedat}
\end{aligned}
\end{equation}
where we have chosen to define the pressure-like functions $P$, $P_2$ such that $P$ coincides with the thermodynamic pressure and $\lambda_3=0$ is a nongeneral artifact from the truncation scheme. From this equation, one could na\"ively expect the transition (for $\Gamma, D_r>0$) to be located at $U\equiv 0$. However, this does not account for the influence of density fluctuations~\cite{caussin2014,*chate2015,*solon2015} and neglects the importance of higher-order correlations at the transition~\cite{kuersten2021}. The actual transition does not occur at the change of sign of this parameter but through a long-wavelength instability of the  homogeneous state. As we have neglected higher order spatial derivatives here for the sake of brevity, these spatial fluctuations as well as viscosity-like terms in the hydrodynamic equation are missing. Including the spatial derivatives by only enslaving locally is conceptually straight-forward. Additionally, allowing fluctuations of the fast modes would lead to a fluctuating force in the final hydrodynamic equation. It is relevant to note that the method used here to close the mode equation hierarchy is only viable because we assume a finite noise strength $D_r$ and does not extend to the deterministic limit. The quantities $\alpha$, $\gamma$ have mean-field and second-order contributions,
\begin{align}
\alpha&=\rho(\frac{\pi}{2} R^2 \Gamma - \frac{128}{90\pi} R^3 v_0^{-1} \Gamma^2) -D_r   \\
\gamma&= \frac{1}{4 D_r} \left( \pi^2 R^2\Gamma + R^3 v_0^{-1} \Gamma^2\frac{256}{45}\right) \text{.}
\end{align}
Already in this minimal demonstration it is apparent, that having access to the Boltzmann equation contributions to second order is imperative for quantitatively meaningful ($v_0$-dependent coefficients) derivation of hydrodynamic equations. This is only reinforced if one strives to go beyond the Toner-Tu paradigm.
\end{document}